\documentclass[preprint,superscriptaddress,amsmath,amssymb,prd,aps,showpacs,floatfix,nofootinbib]{revtex4-1}
\usepackage{graphicx}
\usepackage{dcolumn}
\usepackage{bm}
\usepackage{mathrsfs}
\usepackage{hyperref}
\usepackage{amsmath}
\usepackage{amssymb}
\usepackage{bm}
\usepackage{color}
\usepackage{float}
\usepackage{dcolumn}
\usepackage{multirow}
\usepackage{changepage}
\usepackage{enumerate}
\usepackage[normalem]{ulem}
\usepackage[dvipsnames]{xcolor}
\usepackage{braket}
\usepackage{nicefrac}
\usepackage{multirow}
\usepackage{tabularx}

\hypersetup{pdftex,colorlinks=true,linkcolor=blue,citecolor=red,menucolor=black,urlcolor=blue,filecolor=blue}

\begin{document}
\title{The $\bar B_s^0 \to J/\psi \pi^0 \eta$ decay and the $a_0(980)-f_0(980)$ mixing}
\date{\today}

\author{Jia-Ting Li}
\affiliation{Department of Physics, Guangxi Normal University, Guilin 541004, China}

\author{Jia-Xin Lin}
\affiliation{Department of Physics, Guangxi Normal University, Guilin 541004, China}

\author{Gong-Jie Zhang}
\affiliation{Department of Physics, Guangxi Normal University, Guilin 541004, China}

\author{Wei-Hong Liang}
\email{liangwh@gxnu.edu.cn}
\affiliation{Department of Physics, Guangxi Normal University, Guilin 541004, China}
\affiliation{Guangxi Key Laboratory of Nuclear Physics and Technology, Guangxi Normal University, Guilin 541004, China}

\author{E.~Oset}
\email{oset@ific.uv.es}
\affiliation{Department of Physics, Guangxi Normal University, Guilin 541004, China}
\affiliation{Departamento de F\'{\i}sica Te\'orica and IFIC, Centro Mixto Universidad de
Valencia-CSIC Institutos de Investigaci\'on de Paterna, Aptdo.22085,
46071 Valencia, Spain}


\begin{abstract}
We study the $\bar B_s^0 \to J/\psi f_0(980)$ and $\bar B_s^0 \to J/\psi a_0(980)$ reactions,
and pay attention to the different sources of isospin violation and mixing of $f_0(980)$ and $a_0(980)$ resonances
where these resonances are dynamically generated from meson-meson interaction.
We find that the main cause of isospin violation is the isospin breaking in the meson-meson transition $T$ matrices,
and the other source is that the loops involving kaons in the production mechanism do not cancel due to the different masses of the charged and neutral kaons.
We obtain a branching ratio for $a_0(980)$ production of the order of $5 \times 10^{-6}$.
Future experiments can address this problem, and the production rate and shape of the $\pi^0 \eta$ mass distribution will definitely help to better understand the nature of scalar resonances.

\vspace{0.7cm}

\end{abstract}

\maketitle

\section{Introduction}
\label{sec:intro}

\begin{figure}[b]
   \includegraphics[width=0.4\linewidth]{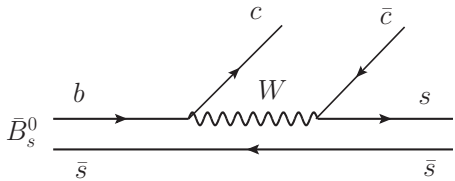}
   \vspace{-0.4cm}
    \caption{Diagram for the $\bar B^0_s$ decay into $J/\psi$ and a primary $s \bar s$ pair. }
    \label{fig:quarklev}
\end{figure}

The $B_s$ decay into $J/\psi$ and two mesons is an excellent source of information on meson dynamics.
At the quark level the decay proceeds via internal emission \cite{chau}, as shown in the diagram of Fig. \ref{fig:quarklev}.
The $c \bar c$ quarks give rise to the $J/\psi$ and the extra $s \bar s$,
which appear in the Cabibbo favored decay mode, have isospin $I=0$.
It is a rather clean process and indeed,
in the LHCb experiment \cite{lhcbexp} the $f_0(980)$ resonance was seen as a strong peak in the invariant mass distribution of $\pi^+ \pi^-$.
The way the $\pi^+ \pi^-$ are produced is studied in detail in Ref.~\cite{baby}.
The $s \bar s$ pair of quarks is hadronized, introducing a $\bar q q$ pair with vacuum quantum numbers,
and $K \bar K$ in $I=0$ plus $\eta \eta$ are produced,
which are allowed to interact within the chiral unitary approach \cite{10us,11us,12us,13us} to produce the $f_0(980)$ resonance,
which is dynamically generated from the interaction of pseudoscalar pairs and couples mostly to $K \bar K$.
With such a clean process producing $I=0$, one finds a very interesting place to produce the $a_0(980)$,
via isospin violation, and add extra information to the subject of the $f_0(980)-a_0(980)$ mixing that has stimulated much work.
Indeed, there are many works devoted to this subject \cite{2wang,3wang,4wang,5wang,6wang,7wang,8wang,9wang,10wang,11wang,
12wang,13wang,14wang,15wang,16wang,17wang,18wang,19wang,20wang,21wang,22wang,23wang,24wang,25wang,vinibayar,aliev,chengli,
liangchen,chengyu} and some cases where, due to a triangle singularity,
the amount of isospin breaking (we prefer this language than the mixing,
since there is not a universal mechanism for the mixing and it depends upon the particular reaction) is abnormally large \cite{beseta,20wang,21wang}.
The way the $a_0(980)$ can be produced in the $B_s \to J/\psi \pi^0 \eta$ decay is tied to the nature of the $f_0(980)$ and $a_0(980)$,
since the resonances are dynamically generated by the pseudoscalar-pseudoscalar ($PP$) interaction \cite{10us}.
It is the meson-meson loops in the Bethe-Salpeter equation,
particularly $K \bar K$ in the case of the $f_0$ and $a_0$, that give rise to the resonances.
The $K^+ K^-$ and $K^0 \bar K^0$ loops cancel for $I=1$ starting from the $I=0$ combination of the hadronized $s \bar s$ quarks,
but only if the masses of the $K^+$ and $K^0$ are taken equal.
When the mass difference is considered, then the isospin is automatically broken
and some peaks appear for the isospin violating decay modes which are rather narrow and are tied to the kaon mass differences.
The relation of the $a_0-f_0$ mixing to this mass difference is shared by most theoretical studies, starting from Ref.~\cite{2wang}.
However, as shown in Ref.~\cite{JiaXinEPJC} in the study of the $D_s \to e^+ \nu_e a_0(980)$,
isospin breaking takes place in the loop for the $K \bar K$ propagation in the decay but also in the same meson-meson scattering matrix,
which enters the evaluation of the process, something already noticed in Ref.~\cite{cpc}.
Yet, the two sources of isospin violation are different depending on the reaction studied, hence
the importance of studying the isospin violation in different processes to gain information
on the way the violation is produced and its dependence on the nature of the $a_0(980)$ and $f_0(980)$ resonances,
which has originated much debate in the literature.

We shall study the process of $a_0$ and $f_0$ production, following the lines of Refs. \cite{baby} and \cite{JiaXinEPJC},
and by taking experimental information on the $B_s \to J/\psi \pi^+ \pi^-$ reaction,
we shall make predictions for the rate of  $B_s \to J/\psi \pi^0 \eta$ production and the shape on the $\pi^0 \eta$ mass distribution.
The branching fraction obtained for this latter decay is of the order of $5\times 10^{-6}$,
well within the range of rates already measured and reported in the PDG \cite{pdg},
which should stimulate its measurement in the future.

\section{Formalism}
\label{sec:form}

The mechanism at the quark level for the $\bar{B}_s^0 \to J/\psi \pi^+ \pi^- (\pi^0 \eta)$ reaction is depicted in Fig.~\ref{fig:quarklev},
having an $s\bar{s}$ pair with isospin $I=0$ at the end.
Note that the light scalars $f_0(980)$ and $a_0(980)$ have $I=0, 1$, respectively.
The production of $f_0(980)$ is isospin conserved,
while the production of $a_0(980)$ is isospin forbidden and involves isospin violation.

To obtain $\pi^+\pi^-$ or $\pi^0\eta$ in the final state in Fig.~\ref{fig:quarklev},
we need to hadronize the $s\bar s$ pair by introducing an extra $\bar q q$ pair with vacuum quantum numbers.
We start with the $q \bar q$ matrix $M$ in SU(3),
\begin{equation}
  \label{eq:M}
  M=\left(\begin{array}
    {ccc}
    u\bar{u} & u\bar{d} & u\bar{s}\\[1mm]
    d\bar{u} & d\bar{d} & d\bar{s}\\[1mm]
    s\bar{u} & s\bar{d} & s\bar{s}
  \end{array}\right).
\end{equation}

Next, we write the matrix $M$ in terms of pseudoscalar mesons,
assuming that the $\eta$ is $\eta_8$ of SU(3),
\begin{equation}\label{eq:PNomix}
  M \to {\mathcal{P}} = \left(
             \begin{array}{ccc}
               \frac{1}{\sqrt{2}}\pi^0 + \frac{1}{\sqrt{6}}\eta + \frac{1}{\sqrt{3}}\eta' & \pi^+ & K^+ \\[2mm]
               \pi^- & -\frac{1}{\sqrt{2}}\pi^0 + \frac{1}{\sqrt{6}}\eta + \frac{1}{\sqrt{3}}\eta' & K^0 \\[2mm]
              K^- & \bar{K}^0 & -\sqrt{\frac{2}{3}}\eta + \sqrt{\frac{1}{3}}\eta' \\
             \end{array}
           \right),
  \end{equation}
which is often used in chiral perturbation theory \cite{10us}.
On the other hand, when we consider the Bramon $\eta-\eta'$ mixing \cite{bramon}, the matrix $M$ can be written as
\begin{equation}\label{eq:Pmix}
  M \to {\mathcal{P}}^{(\mathrm{m})} =
           \left(
             \begin{array}{ccc}
               \frac{1}{\sqrt{2}}\pi^0 + \frac{1}{\sqrt{3}}\eta + \frac{1}{\sqrt{6}}\eta' & \pi^+ & K^+ \\[2mm]
               \pi^- & -\frac{1}{\sqrt{2}}\pi^0 + \frac{1}{\sqrt{3}}\eta + \frac{1}{\sqrt{6}}\eta' & K^0 \\[2mm]
              K^- & \bar{K}^0 & -\frac{1}{\sqrt{3}}\eta + \sqrt{\frac{2}{3}}\eta' \\
             \end{array}
           \right).
  \end{equation}
Since the $\eta'$ is inessential in the dynamical generation of the $f_0(980)$ and $a_0(980)$ resonances \cite{10us},
we will ignore the $\eta'$ in the present work.

After hadronization of the $s \bar s$ component, we obtain
\begin{equation}\label{eq:H1}
  s\bar s \to H = \sum_i s\, \bar q_i q_i \, \bar s = \sum_i {\mathcal{P}}_{3i} \; {\mathcal{P}}_{i3}\, = ({\mathcal{P}}^2)_{33}.
\end{equation}
In the case without $\eta-\eta'$ mixing, the matrix $\mathcal{P}$ of Eq.~\eqref{eq:PNomix} is used,
and then the hadron component $H$ in Eq.~\eqref{eq:H1} is given by
\begin{equation}\label{eq:HNomix}
  H= K^-K^+ + \bar K^0 K^0 + \frac{2}{3}\, \eta \eta.
\end{equation}

In the case with $\eta-\eta'$ mixing, one uses matrix $\mathcal{P}^{(m)}$ of Eq.~\eqref{eq:Pmix},
and obtains
\begin{equation}\label{eq:Hmix}
  H= K^-K^+ + \bar K^0 K^0 + \frac{1}{3}\, \eta \eta,
\end{equation}
differing only in the $\eta\eta$ component,
which affects the production of $f_0$ but not the production of $a_0$.
We define the weight of the $PP$ components in $H$ as
\begin{equation}\label{eq:weight}
  h_{K^+K^-}=1,~~~~ h_{K^0 \bar K^0}=1,~~~~ h_{\eta\eta}=\frac{2}{3},~~~~ h_{\eta\eta}^{(\mathrm{m})}=\frac{1}{3}.
\end{equation}

\begin{figure}[t]
  \includegraphics[width=1\linewidth]{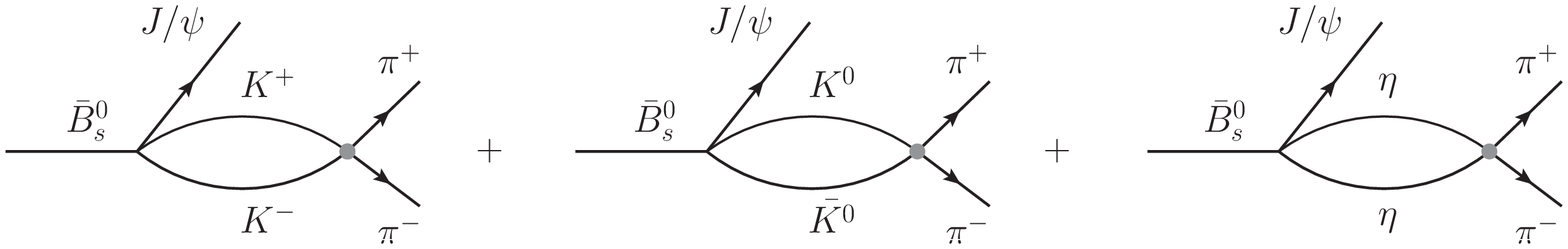}\\[0.2cm]
   \includegraphics[width=1\linewidth]{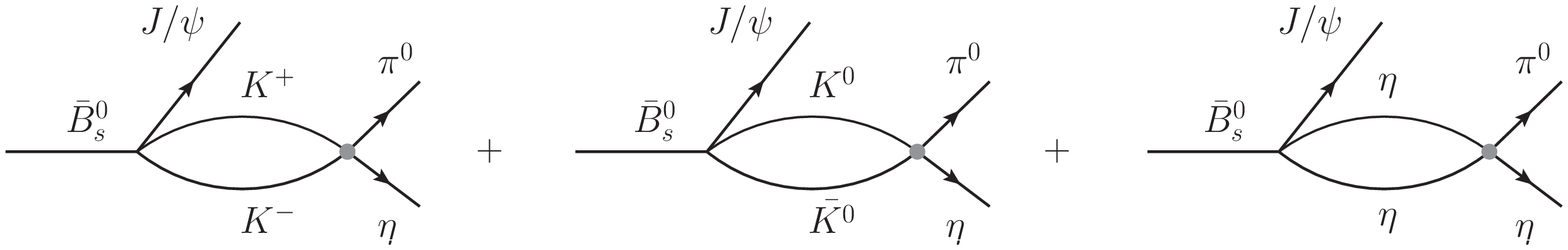}
   \caption{Final state interaction of the hadron components leading to $\pi^+\pi^-$ or $\pi^0 \eta$ in the final state.}
   \label{fig:FeynDiag}
\end{figure}

One can see that neither Eq.~\eqref{eq:HNomix} nor Eq.~\eqref{eq:Hmix} contains $\pi^+ \pi^-$ or $\pi^0 \eta$,
but they can be produced by the final state interaction of the $K \bar K$ and $\eta \eta$ components,
as depicted in Fig.~\ref{fig:FeynDiag}.
The transition matrix from the $PP$ state to $\pi^+ \pi^-$ or $\pi^0 \eta$
is represented by the circle behind the meson-meson loop in Fig.~\ref{fig:FeynDiag},
which contains the information of the $f_0(980)$ and $a_0(980)$ respectively.
According to the method in Ref.~\cite{10us} (the chiral unitary approach), these resonances are the result of the $PP$ interaction in the coupled channels $K\bar K, \pi\pi, \pi\eta, \eta\eta$.

By using the unitary normalization \cite{10us,JiaXinEPJC},
the amplitude for the $\bar{B}_s^0 \to J/\psi \pi^+ \pi^-$ decay,
as a function of the $\pi^+\pi^-$ invariant mass $M_{\rm inv}(\pi^+\pi^-)$,
is given by \cite{JiaXinEPJC}
\begin{eqnarray}\label{eq:tpipi}
  t_{\pi^+\pi^-} &=& {\mathcal{C}}\left[ h_{K^+K^-} \cdot G_{K^+K^-} (M_{\rm inv}(\pi^+\pi^-)) \cdot T_{K^+K^-,\pi^+\pi^-}(M_{\rm inv}(\pi^+\pi^-))
  \right.\nonumber \\
  &&~ + h_{K^0 \bar K^0} \cdot G_{K^0 \bar K^0} (M_{\rm inv}(\pi^+\pi^-)) \cdot T_{K^0 \bar K^0,\pi^+\pi^-}(M_{\rm inv}(\pi^+\pi^-))\nonumber \\
  &&~ + h_{\eta\eta} \times 2\times \frac{1}{2} \cdot \left. G_{\eta\eta} (M_{\rm inv}(\pi^+\pi^-)) \cdot T_{\eta\eta,\pi^+\pi^-}(M_{\rm inv}(\pi^+\pi^-))\right],
  \end{eqnarray}
and the amplitude for the $\bar{B}_s^0 \to J/\psi \pi^0 \eta$  decay,
as a function of the $\pi^0\eta$ invariant mass $M_{\rm inv}(\pi^0\eta)$,
is given by \cite{JiaXinEPJC}
\begin{eqnarray}\label{eq:tpieta}
  t_{\pi^0\eta} &=& {\mathcal{C}}\left[ h_{K^+K^-} \cdot G_{K^+K^-} (M_{\rm inv}(\pi^0\eta)) \cdot T_{K^+K^-,\pi^0\eta}(M_{\rm inv}(\pi^0\eta))
  \right.\nonumber \\
  &&~ + h_{K^0 \bar K^0} \cdot G_{K^0 \bar K^0} (M_{\rm inv}(\pi^0\eta)) \cdot T_{K^0 \bar K^0,\pi^0\eta}(M_{\rm inv}(\pi^0\eta))\nonumber \\
  &&~ + h_{\eta\eta} \times 2\times \frac{1}{2} \cdot \left. G_{\eta\eta} (M_{\rm inv}(\pi^0\eta)) \cdot T_{\eta\eta,\pi^0\eta}(M_{\rm inv}(\pi^0\eta))\right],
  \end{eqnarray}
with $\mathcal{C}$ an arbitrary normalization constant which cancels in the ratio of the $f_0$, $a_0$ production rates.
For the case with $\eta-\eta'$ mixing, the corresponding amplitudes can be obtained
by replacing $h_{\eta\eta}$ with $h_{\eta\eta}^{(m)}$ in Eqs.~\eqref{eq:tpipi} and \eqref{eq:tpieta}.

In Eqs.~\eqref{eq:tpipi} and \eqref{eq:tpieta},
$G_i$ is the loop function of the two intermediate pseudoscalar mesons,
which is regularized with a three momentum cut-off $q_{\rm max}$ \cite{10us},
\begin{equation}\label{eq:Gfuction}
  G_i(\sqrt{s})=\int_0^{q_{\rm max}} \frac{q^2\; {\rm d}q}{(2\pi)^2}\; \frac{w_1+w_2}{w_1\, w_2 \,[s-(w_1 +w_2)^2+i \epsilon]},
\end{equation}
with $w_j=\sqrt{m_j^2+\vec q^{\,2}}$ and $\sqrt{s}$ the centre-of-mass energy of the two mesons in the loop.
$T_{i,j}$ is the total amplitude for the $i\to j$ transition and can be obtained
by solving the Bethe-Salpeter (BS) equation with six $PP$ coupled channels $\pi^+\pi^-$, $\pi^0\pi^0$, $K^+K^-$, $K^0\bar{K}^0$, $\eta\eta$ and $\pi^0\eta$, in a matrix form,
\begin{equation}\label{eq:BSeq}
  T=[1-V\,G]^{-1}\, V,
\end{equation}
where the matrix $V$ is the kernel of BS equation.
Its elements $V_{ij}$ are the $s$-wave transition potentials which can be taken from Eq.~(A3) and Eq.(A4) of Ref.~\cite{JiaXinEPJC},
corresponding to the cases without and with $\eta-\eta'$ mixing, respectively.

The differential decay width for $\bar B_s^0\to J/\psi\pi^0\eta$ or $\bar B_s^0\to J/\psi\pi^+\pi^-$
decay is given by
\begin{eqnarray}\label{eq:dGam}
  \frac{{\rm d} \Gamma}{{\rm d} M_{\rm inv}(ij)}=\frac{1}{(2\pi)^3} \;
  \frac{1}{4M_{\bar B_s^0}^2}\; \frac{1}{3}\; p_{J/\psi}^2 \; p_{J/\psi}\; \tilde{p}_{\pi}\; |t_{ij}|^2,
  \end{eqnarray}
where $ij=\pi^+ \pi^-$ or $\pi^0 \eta$,
$M_{\rm inv}(ij)$ is the invariant mass of the final $\pi^+ \pi^-$ or $\pi^0 \eta$,
$t_{\pi^+ \pi^-}$ and $t_{\pi^0 \eta}$ are the amplitudes from Eq.~\eqref{eq:tpipi} and Eq.~\eqref{eq:tpieta} respectively,
$p_{J/\psi}$ is the $J/\psi$ momentum in the $\bar B_s^0$ rest frame,
and $\tilde{p}_{\pi}$ is the pion momentum in the rest frame of the $\pi^+ \pi^-$ or $\pi^0 \eta$ system,
\begin{eqnarray}\label{eq:pjpsi}
  p_{J/\psi}=\frac{\lambda^{1/2}(M^2_{\bar B^0_s},M^2_{J/\psi},M_{\rm inv}^2)}{2M_{\bar B^0_s}},
\end{eqnarray}
\begin{align}\label{eq:ppi}
  \tilde{p}_{\pi}=%
      \begin{cases}
          \dfrac{\lambda^{1/2}(M^2_{\rm inv},m^2_{\pi},m^2_{\pi})}{2M_{\rm inv}}, & {\rm for~\pi^+\pi^-~production,} \\[0.5cm]
          \dfrac{\lambda^{1/2}(M^2_{\rm inv},m^2_{\pi},m^2_{\eta})}{2M_{\rm inv}}, & {\rm for~\pi^0\eta~production,}
      \end{cases}
\end{align}%
with $\lambda(x^2, y^2, z^2)=x^2+y^2+z^2-2xy-2yz-2zx$ the K\"allen function.
In Eq.~\eqref{eq:dGam}, the factor $\frac{1}{3}\, p_{J/\psi}^2$ stems from the fact that we need a $p$-wave to match angular momentum in the $0^- \to 1^-\, 0^+$ transition and we take a vertex of type $p_{J/\psi}\, \cos \theta$.

\section{Results}
\label{sec:result}
\begin{figure}[t]
  \begin{center}
  \includegraphics[scale=0.75]{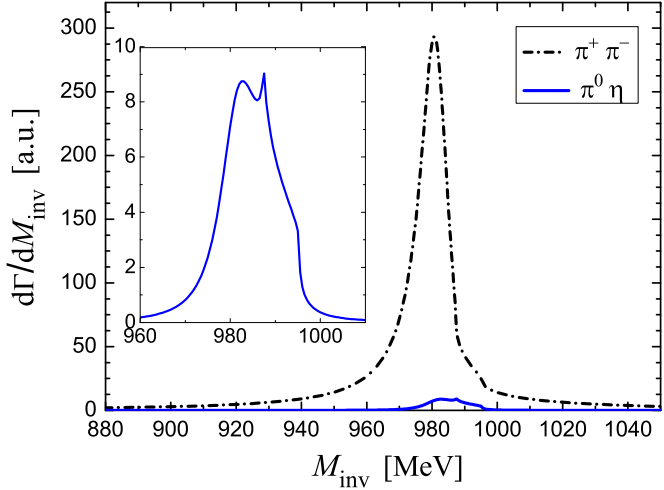}
  \end{center}
  \vspace{-1cm}
  \caption{$M_{\rm inv}(\pi^+ \pi^-)$ mass distribution for $\bar B_s^0  \to J/\psi f_0(980), f_0(980) \to \pi^+ \pi^-$ decay, and $M_{\rm inv}(\pi^0 \eta)$ mass distribution for $\bar B_s^0 \to J/\psi a_0(980), a_0(980) \to \pi^0 \eta$ decay. Inset: Magnified $\pi^0 \eta$. (Without $\eta-\eta'$ mixing)}
  \label{Fig:MinvNoMix}
  \end{figure}

\begin{figure}[t]
  \begin{center}
  \includegraphics[scale=0.75]{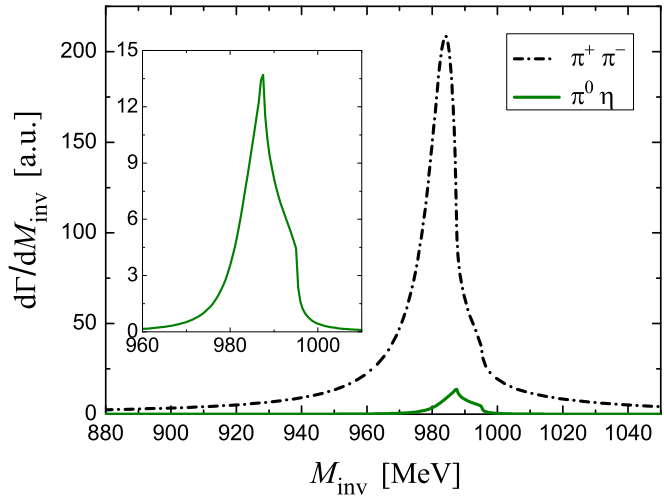}
  \end{center}
  \vspace{-1cm}
  \caption{$M_{\rm inv}(\pi^+ \pi^-)$ mass distribution for $\bar B_s^0  \to J/\psi f_0(980), f_0(980) \to \pi^+ \pi^-$ decay, and $M_{\rm inv}(\pi^0 \eta)$ mass distribution for $\bar B_s^0  \to J/\psi a_0(980), a_0(980) \to \pi^0 \eta$ decay. Inset: Magnified $\pi^0 \eta$. (With $\eta-\eta'$ mixing)}
  \label{Fig:MinvMix}
  \end{figure}

We follow Ref.~\cite{JiaXinEPJC} and take the cut-off $q_{\rm max}=600$ MeV and $650$ MeV
for the cases without $\eta-\eta'$ mixing and with $\eta-\eta'$ mixing respectively,
with which the $f_0(980)$ and $a_0(980)$ resonances can be dynamically produced well from the PP interaction.
The $\pi^+ \pi^-$ and $\pi^0 \eta$ mass distributions $\frac{{\rm d} \Gamma}{{\rm d} M_{\rm inv}(ij)}$ are shown in Fig.~\ref{Fig:MinvNoMix} for the case without $\eta-\eta'$ mixing and in Fig.~\ref{Fig:MinvMix} for the case with $\eta-\eta'$ mixing, respectively.
By comparing Fig.~\ref{Fig:MinvNoMix} and Fig.~\ref{Fig:MinvMix},
one finds that the results of the two figures are very similar,
and the difference between them can serve as an estimate of the uncertainties of our formalism.

Now, let us look at the $\pi^+ \pi^-$ and $\pi^0 \eta$ mass distributions in Fig.~\ref{Fig:MinvMix} with $\eta-\eta'$ mixing.
One can see a strong peak for $f_0(980)$ production in the $\pi^+ \pi^-$ mass distribution
and a small peak for $a_0(980)$ production in the $\pi^0 \eta$ mass distribution.
Here the shape of $a_0(980)$ resonance is quite narrow,
much different to the standard cusp-like shape (with a width of about 120 MeV) of the ordinary production of $a_0(980)$ in an isospin allowed reaction \cite{BESIII:2016tqo}.
If isospin were conserved, one would find the $a_0(980)$ production with zero strength.
The small peak of $a_0(980)$ in Fig.~\ref{Fig:MinvMix} indicates that isospin violation takes places in the
$\bar{B}_s^0 \to J/\psi \pi^0 \eta$ reaction.
According to Eq.~(A4) of Ref.~\cite{JiaXinEPJC}, we have $V_{K^+K^-,\pi^0\eta}=-V_{K^0 \bar K^0,\pi^0\eta}$ for the transition potentials.
Hence, if we use average masses for kaons,
there will be a precise cancellation of the first two terms of the amplitude $t_{\pi^0\eta}$ in Eq.~\eqref{eq:tpieta},
resulting on zero strength for $a_0(980)$ production.
On the contrary,
using the physical masses for the neutral $K^0$ and the charged $K^+$ in the formalism will result
in the production of the $a_0(980)$ resonance with a narrow shape related to the difference of mass between the charged and neutral kaons.
In our picture, there are two sources of isospin violation:
one is the $K^+$, $K^0$ mass difference for the explicit $K^+ K^-$ and $K^0 \bar K^0$ loops in Fig.~\ref{fig:FeynDiag},
the other one is from the $T$ matrix involving rescattering in Fig.~\ref{fig:FeynDiag}.

It will be interesting to see the effects of the two sources of isospin violation.
For that, we follow Ref.~\cite{JiaXinEPJC} and define the ratio $R$,
which reflects the amount of the isospin violation, as
\begin{equation}\label{eq:R}
  R=\frac{\Gamma(\bar B_s^0 \to J/\psi a_0(980), a_0(980) \to \pi^0 \eta )}
        {\Gamma(\bar B_s^0 \to J/\psi f_0(980), f_0(980) \to \pi^+ \pi^-)},
\end{equation}
with decay widths $\Gamma[\bar B_s^0 \to J/\psi a_0(980), a_0(980) \to \pi^0 \eta ]$ and $\Gamma[\bar B_s^0 \to J/\psi f_0(980), f_0(980) \to \pi^+ \pi^-]$ obtained by integrating Eq.~\eqref{eq:dGam} over the invariant mass $M_{\rm inv}(ij)$.

Under several different assumptions related to the two sources of isospin violation,
we evaluate the ratio $R$. The results are shown in Table \ref{tab:R}.
\begin{table*}[b]
\renewcommand\arraystretch{1.1}
\centering
\caption{\vadjust{\vspace{-2pt}}values of $R$ with different assumptions. (In the table, I.V. denotes isospin violation.)}\label{tab:R}
\begin{tabular}{l|l|c}
\hline \hline
\multirow{3}{*}{no $\eta-\eta'$ mixing}     & ~I.V. both in $T$ matrix and in explicit $K\bar K$ loops (Case 1) &$3.1\times 10^{-2}$\\
\cline{2-3}
                                            & ~I.V. only in $T$ matrix (Case 2)& ~~$3.5\times 10^{-2}$~~\\
\cline{2-3}
                                            & ~I.V. only in explicit $K\bar K$ loops (Case 3)& $7.0\times 10^{-4}$ \\
\hline\hline
\multirow{3}{*}{with $\eta-\eta'$ mixing}~  & ~I.V. both in $T$ matrix and in explicit $K\bar K$ loops (Case 4)& $3.7\times 10^{-2}$ \\
\cline{2-3}
                                            &~I.V. only in $T$ matrix (Case 5) & $4.1\times 10^{-2}$\\
\cline{2-3}
                                            & ~I.V. only in explicit $K\bar K$ loops (Case 6)& $9.7\times 10^{-4}$ \\
\hline\hline
\end{tabular}
\end{table*}

From Table \ref{tab:R},
we observe that the ratio $R$ with $\eta-\eta'$ mixing (Case 4) is about $20\%$ bigger
than that without $\eta-\eta'$ mixing (Case 1).
By comparing the values of $R$ for Case 2 and Case 3 (or, for Case 5 and Case 6),
we find that the isospin violation in the $T$ matrix has more important effect than that in the explicit $K\bar K$ loops, being at least one order of magnitude larger.
This fact is interesting, since in our picture the $f_0(980)$ and $a_0(980)$ resonances are dynamically generated from the $PP$ interaction with the information on their nature contained in the $T$ matrix.
For the $\bar B_s^0 \to J/\psi \pi^+ \pi^- (\pi^0 \eta)$ decay, neither the $\pi^+ \pi^-$ nor the $\pi^0 \eta$ can be directly produced from the $s\bar s$ hadronization [see Eqs.~\eqref{eq:HNomix} and \eqref{eq:Hmix}],
hence there is no contribution from the tree level. Instead, they are produced through the rescattering mechanism of Fig.~\ref{fig:FeynDiag},
with $f_0(980)$ and $a_0(980)$ resonances as dynamically generated states from the $PP$ interaction.
The production rate of the $f_0(980)$, ($a_0(980)$) resonances in the $\bar B_s^0 \to J/\psi \pi^+ \pi^- (\pi^0 \eta)$ decay
is sensitive to the resonance information contained by the $T$ matrix.
Therefore, this mode is particularly suitable to test the nature of $f_0(980)$ and $a_0(980)$ resonances
and to investigate the isospin violation.

From PDG \cite{pdg}, the experimental branching ratio of the $\bar B_s^0 \to J/\psi f_0(980), f_0(980) \to \pi^+ \pi^-$ decay reads
\begin{equation}\label{eq:Br}
  {\mathrm{Br}}[\bar B_s^0 \to J/\psi f_0(980), f_0(980) \to \pi^+ \pi^-]=(1.28 \pm 0.18)\times 10^{-4}.
\end{equation}
By using the ratio $R$ in Table ~\ref{tab:R} and the branching ratio of Eq.~\eqref{eq:Br},
the branching ratio for $a_0(980)$ production can be obtained,
\begin{align}\label{eq:Bra0}
  {\mathrm{Br}}[\bar B_s^0 \to J/\psi a_0(980), a_0(980) \to \pi^0 \eta]=%
      \begin{cases}
          (3.95\pm 0.56)\times 10^{-6}, & {\rm for~ Case~1;} \\[0.1cm]
          (4.74\pm 0.67)\times 10^{-6}, & {\rm for~ Case~4.}
      \end{cases}
\end{align}%
This branching ratio is the order of $5\times 10^{-6}$,
not too small considering that several rates of the order of $10^{-7}$ are tabulated in the PDG \cite{pdg}.
The branching ratio and the shape of the $\pi^0 \eta$ mass distribution of the $\bar B_s^0 \to J/\psi \pi^0 \eta$ decay provide relevant information on the nature of the $a_0(980)$ resonance. Experimental measurements will be very valuable.

\section{Conclusions}
\label{sec:concl}
In the present work, we study the isospin allowed decay process $\bar B_s^0 \to J/\psi \pi^+ \pi^-$
and the isospin forbidden decay process $\bar B_s^0 \to J/\psi \pi^0 \eta$,
paying attention to the different sources of isospin violation.

First, we have $J/\psi~s\bar s$ production in the $\bar B_s^0$ decay, via internal emission of Fig.~\ref{fig:quarklev}.
After the hadronization of $s \bar s$ into meson-meson components,
we obtain the $K\bar K$ pairs and $\eta \eta$, while $\pi^+ \pi^-$ and $\pi^0 \eta$ are not produced at this step.
Therefore, to see $\pi^+\pi^-$ or $\pi^0\eta$ in the final state,
the rescattering of the $K \bar K$, $\eta \eta$ components is needed to produce $\pi^+ \pi^-$ and $\pi^0 \eta$ at the end.
The picture shows that the weak decay amplitudes are proportional to the $T$ matrix of the meson-meson transitions.
We can obtain the information about the violation of isospin from these magnitudes.
In Figs.~\ref{Fig:MinvNoMix} and \ref{Fig:MinvMix},
we observe a clear signal for $f_0(980)$ production.
We also observe that the shape of the $\pi^0 \eta$ mass distribution is very different from the shape of the common $a_0(980)$ production in the isospin allowed reactions,
and it is related to the difference in mass between the charged and neutral kaons.
In the production of $a_0(980)$
we find two sources of isospin violation: one is that the loops containing $K^+ K^-$ or $K^0 \bar K^0$ do not cancel due to the different mass between the charged and neutral kaons,
and the other is that the transition $T$ matrix of meson-meson interaction already contains some isospin violation.
In fact, we find that the contribution from isospin violation in the $T$ matrix is far more important than the contribution of the explicit loops in the weak decay,
being at least one order of magnitude larger.
The study done here shows that this reaction is very sensitive to the way the resonances are generated.

The $D_s^+$ semileptonic decay \cite{JiaXinEPJC} and $\bar B_s^0$ mesonic decay both get an $s \bar s$ pair at the end,
and the two resonances of $f_0(980)$ and $a_0(980)$ are produced dynamically by the interaction of pseudoscalar mesons through the chiral unitary approach.
The results of $D_s^+$ semileptonic decay are consistent with the experimental upper bound.
We also calculate the branching ratio of $\bar B_s^0 \to J/\psi a_0(980)$ for $a_0(980)$ production,
and the values are not too small, of the order of $5\times 10^{-6}$.
Our results provide a reference basis for the experiment,
which we expect to be carried out in the near future.

\begin{acknowledgments}
This work is  partly supported by the National Natural Science Foundation of China under Grants No. 11975083 and No. 12147211.
This work is also partly supported by the Spanish Ministerio de Economia y Competitividad
and European FEDER funds under Contracts No. FIS2017-84038-C2-1-P B
and by Generalitat Valenciana under contract PROMETEO/2020/023.
This project has received funding from the European Unions Horizon 2020 research and innovation programme
under grant agreement No. 824093 for the ``STRONG-2020" project.
\end{acknowledgments}



\begin{thebibliography}{}

\bibitem{chau}
L.~L.~Chau,
Phys. Rept. \textbf{95}, 1-94 (1983).

\bibitem{lhcbexp}
R.~Aaij \textit{et al.} [LHCb],
Phys. Lett. B \textbf{698}, 115-122 (2011).

\bibitem{baby}
W.~H.~Liang and E.~Oset,
Phys. Lett. B \textbf{737}, 70-74 (2014)

\bibitem{10us}
J.~A.~Oller and E.~Oset,
Nucl. Phys. A \textbf{620}, 438-456 (1997);
[Erratum: Nucl. Phys. A \textbf{652}, 407-409 (1999)].

\bibitem{11us}
N.~Kaiser,
Eur. Phys. J. A \textbf{3}, 307-309 (1998).

\bibitem{12us}
M.~P.~Locher, V.~E.~Markushin and H.~Q.~Zheng,
Eur. Phys. J. C \textbf{4}, 317-326 (1998).

\bibitem{13us}
J.~Nieves and E.~Ruiz Arriola,
Nucl. Phys. A \textbf{679}, 57-117 (2000).

\bibitem{2wang}
N.~N.~Achasov, S.~A.~Devyanin and G.~N.~Shestakov,
Phys. Lett. B \textbf{88}, 367-371 (1979).

\bibitem{3wang}
N.~N.~Achasov, S.~A.~Devyanin and G.~N.~Shestakov,
Yad. Fiz. \textbf{33}, 1337-1348 (1981); Sov. J. Nucl. Phys. \textbf{33}, 715 (1981).

\bibitem{4wang}
N.~N.~Achasov and G.~N.~Shestakov,
Phys. Rev. D \textbf{56}, 212-220 (1997).

\bibitem{5wang}
O.~Krehl, R.~Rapp and J.~Speth,
Phys. Lett. B \textbf{390}, 23-28 (1997).

\bibitem{6wang}
B.~Kerbikov and F.~Tabakin,
Phys. Rev. C \textbf{62}, 064601 (2000).

\bibitem{7wang}
F.~E.~Close and A.~Kirk,
Phys. Lett. B \textbf{489}, 24-28 (2000).

\bibitem{8wang}
A.~E.~Kudryavtsev and V.~E.~Tarasov,
JETP Lett. \textbf{72}, 410-414 (2000); Pisma Zh. Eksp. Teor. Fiz. \textbf{72}, 589-594 (2000).

\bibitem{9wang}
V.~Y.~Grishina, L.~A.~Kondratyuk, M.~Buescher, W.~Cassing and H.~Stroher,
Phys. Lett. B \textbf{521}, 217-224 (2001).

\bibitem{10wang}
F.~E.~Close and A.~Kirk,
Phys. Lett. B \textbf{515}, 13-16 (2001)

\bibitem{11wang}
A.~E.~Kudryavtsev, V.~E.~Tarasov, J.~Haidenbauer, C.~Hanhart and J.~Speth,
Phys. Rev. C \textbf{66}, 015207 (2002).

\bibitem{12wang}
L.~A.~Kondratyuk, E.~L.~Bratkovskaya, V.~Y.~Grishina, M.~Buescher, W.~Cassing and H.~Stroher,
Phys. Atom. Nucl. \textbf{66}, 152-171 (2003); Yad. Fiz. \textbf{66}, 155-174 (2003).

\bibitem{13wang}
N.~N.~Achasov and A.~V.~Kiselev,
Phys. Lett. B \textbf{534}, 83-86 (2002).

\bibitem{14wang}
N.~N.~Achasov and G.~N.~Shestakov,
Phys. Rev. Lett. \textbf{92}, 182001 (2004).

\bibitem{15wang}
V.~Y.~Grishina, L.~A.~Kondratyuk, M.~Buescher and W.~Cassing,
Eur. Phys. J. A \textbf{21}, 507-520 (2004).

\bibitem{16wang}
N.~N.~Achasov and G.~N.~Shestakov,
Phys. Rev. D \textbf{70}, 074015 (2004).

\bibitem{17wang}
J.~J.~Wu, Q.~Zhao and B.~S.~Zou,
Phys. Rev. D \textbf{75}, 114012 (2007).

\bibitem{18wang}
J.~J.~Wu and B.~S.~Zou,
Phys. Rev. D \textbf{78}, 074017 (2008).

\bibitem{19wang}
C.~Hanhart, B.~Kubis and J.~R.~Pelaez,
Phys. Rev. D \textbf{76}, 074028 (2007).

\bibitem{20wang}
J.~J.~Wu, X.~H.~Liu, Q.~Zhao and B.~S.~Zou,
Phys. Rev. Lett. \textbf{108}, 081803 (2012).

\bibitem{21wang}
F.~Aceti, W.~H.~Liang, E.~Oset, J.~J.~Wu and B.~S.~Zou,
Phys. Rev. D \textbf{86}, 114007 (2012).

\bibitem{22wang}
L.~Roca,
Phys. Rev. D \textbf{88}, 014045 (2013).

\bibitem{23wang}
V.~E.~Tarasov, W.~J.~Briscoe, W.~Gradl, A.~E.~Kudryavtsev and I.~I.~Strakovsky,
Phys. Rev. C \textbf{88}, 035207 (2013).

\bibitem{24wang}
T.~Sekihara and S.~Kumano,
Phys. Rev. D \textbf{92}, 034010 (2015).

\bibitem{25wang}
F.~Aceti, J.~M.~Dias and E.~Oset,
Eur. Phys. J. A \textbf{51}, 48 (2015).

\bibitem{vinibayar}
M.~Bayar and V.~R.~Debastiani,
Phys. Lett. B \textbf{775}, 94-99 (2017).

\bibitem{aliev}
T.~M.~Aliev and S.~Bilmis,
Eur. Phys. J. A \textbf{54}, 147 (2018).

\bibitem{chengli}
X.~D.~Cheng, H.~B.~Li, R.~M.~Wang and M.~Z.~Yang,
Phys. Rev. D \textbf{99}, 014024 (2019).


\bibitem{liangchen}
W.~H.~Liang, H.~X.~Chen, E.~Oset and E.~Wang,
Eur. Phys. J. C \textbf{79}, 411 (2019).

\bibitem{chengyu}
X.~D.~Cheng, R.~M.~Wang and Y.~G.~Xu,
Phys. Rev. D \textbf{102}, 054009 (2020).

\bibitem{beseta}
M.~Ablikim \textit{et al.} [BESIII],
Phys. Rev. Lett. \textbf{108}, 182001 (2012).

\bibitem{JiaXinEPJC}
J.~X.~Lin, J.~T.~Li, S.~J.~Jiang, W.~H.~Liang and E.~Oset,
Eur. Phys. J. C \textbf{81}, 1017 (2021).

\bibitem{cpc}
W.~H.~Liang, S.~Sakai, J.~J.~Xie and E.~Oset,
Chin. Phys. C \textbf{42}, 044101 (2018).

\bibitem{pdg}
P.A.~Zyla \textit{et al.} [Particle Data Group],
Prog. Theor. Exp. Phys. \textbf{2020}, 083C01 (2020) and 2021 update.

\bibitem{bramon}
A.~Bramon, A.~Grau and G.~Pancheri,
Phys. Lett. B \textbf{283}, 416-420 (1992).

\bibitem{BESIII:2016tqo}
M.~Ablikim \textit{et al.} [BESIII],
Phys. Rev. D \textbf{95}, 032002 (2017).

\end{thebibliography}
\end{document}